\documentclass[sn-vancouver,Numbered,iicol]{sn-jnl} 

\usepackage{graphicx} 
\usepackage{multirow} 
\usepackage{amsmath,amssymb,amsfonts} 

\usepackage{alphabeta} 
\usepackage{siunitx}	
\usepackage{orcidlink} 

\begin{document}


\title{Manuscript preparation guide for Radiological~Physics~and~Technology}

\author*[1]{Nobuyuki Kanematsu \orcidlink{0000-0002-2534-9933}} 
\email{kanematsu.nobuyuki@qst.go.jp}

\affil[1]{Department of Accelerator and Medical Physics, Institute for Quantum Medical Science, National Institutes for Quantum Science and Technology, 4-9-1 Anagawa, Inage-ku, Chiba 263-8555, Japan}

\abstract{This guide is intended for authors who wish to publish an original paper in \emph{Radiological Physics and Technology}.
They should follow the submission guidelines of the journal as well as standard guidelines and conventions for scientific writing.
The journal recommends a standard structure for the main text and requires an additional section for declarations.
Each section has a numbered heading that announces the topic of its content.
The text may consist of paragraphs for subtopics.
Original content should be detailed while unoriginal content available elsewhere should be minimized and distinguished by reference citation.
Quantities and equations should be presented according to academic conventions.
Tables and figures should only be used to present essential information clearly and efficiently.
The \LaTeX\ source files of this guide are available online and may be used as a manuscript template.
A checklist is also provided for submitting authors.}


\maketitle


\section*{About this guide} 

This guide is intended for authors who wish to publish an original paper in the journal \emph{Radiological Physics and Technology} (ISSN:1865-0333/0341 for print/electronic, Springer Nature) \cite{RPT-website}.
It generally complies with the submission guidelines of the journal \cite{RPT-guidelines}, the editorial policies for Springer journals \cite{Springer-policies}, and the recommendations by the International Committee of Medical Journal Editors \cite{ICMJE-recommendations}.

This guide was prepared using the free typesetting software \LaTeX\ \cite{LaTeX-Project} with the class file sn-jnl.cls for Springer Nature journals \cite{Springer-latex}.
The \href{https://arxiv.org/pdf/2403.02888.pdf}{article} was posted on the preprint server \emph{arXiv} (Cornell Tech, New York, USA) with a non-exclusive license to distribute and is accessible at \url{https://arxiv.org/abs/2403.02888}.
The \LaTeX\ \href{https://arxiv.org/src/2403.02888}{source files} are available there, and \LaTeX\ users can use them as a manuscript template for submission to the journal with their original content in a single-column, double-spaced, and line-numbered format.

This guide offers practical instructions to authors on how to write a title page, a main text in numbered sections, and a back matter in optional unnumbered sections, using minimum examples of common components.
The references include several online documents last accessed on \today.
Appendix~\ref{appA} explains how to use appendices.
Appendix~\ref{appB} shows an example of files to be submitted, provides a checklist for submitting authors, and illustrates a typical progression of an initial submission. 
Appendix~\ref{appC} summarizes the peer review process.

\section*{General principles} 

\begin{itemize}
\item
Original papers are to transfer new knowledge from research into science accurately, efficiently, and openly.
Clarity, rigorousness, conciseness, and focus on novelties are then essential.
\item
Each section, subsection, or subsubsection has a numbered heading that announces the topic of its content. 
When the text comprises multiple paragraphs, a subtopic for each should be mentioned in the first sentence.
\item
Intended readers are peer researchers who may use the paper for their own research.
They will be represented by the reviewers who volunteer to assess the manuscript and are expected to assist the authors with revisions.
\end{itemize}

\section*{Title page} 

\bmhead{Title}
The title should concisely announce the research topic, purpose, and methodology without using trade names or uncommon abbreviations.
Title words should be reasonably understandable for all readers of the journal.
 
 \bmhead{Authors}
All authors must have contributed to the work for both research and manuscript, have approved the manuscript for submission, and be accountable for all aspects of the work \cite{ICMJE-recommendations}.
The corresponding author should provide an official e-mail address for communication with readers.\footnote{The corresponding author for publication may not necessarily be the same as the corresponding author for submission, who is the submitter.} 
Any changes to the author list after the initial submission of the manuscript must be justified for approval.

\bmhead{Abstract}
The abstract should briefly summarize the main text without figures, tables, equations, or references, and must be self-contained and independent of the main text, and vice versa. 
The purpose, methods, results, and conclusions should be stated without technical details to highlight the essence of the research.
The maximum length is 250 words for a full-length paper or 150 words for a concise paper, depending on the article type.

\bmhead{Keywords}
Select four to six keywords from the technical terms and concepts used in the manuscript.


\section{Introduction} 

The first paragraph of the introduction section typically describes the research topic based on evidence, including its position in the field, relevance to clinical practice, and current status.

The following paragraphs typically identify the problem associated with the topic, discuss its implications, and review the preceding research that has dealt with similar problems or methods relevant to such problems.

The last paragraph of the section typically announces the approach to solve the problem and the objectives to be achieved in the current research.

\section{Materials and methods}

\subsection{General}

The materials-and-methods section should consist of structured subsections and subsubsections to describe the materials used, such as objects, equipment, theory, and data; and the methods used, such as experiment, calculation, analysis, and testing.

Descriptions of the materials and methods must be rigorous and complete to be reproducible by readers, yet as concise as possible.
Naturally, original content should be detailed while unoriginal content available elsewhere should be minimized and distinguished by reference citation.

\subsection{Reference citation}
In the context of arguments, non-obvious facts must be cited from the minimum necessary references.
The journal follows the National Library of Medicine or Vancouver referencing style \cite{NLM-style, NLM-samples}, where citations are sequentially numbered in the main text for index listing of the references in the back matter.

\subsection{Quantities, symbols, units, and equations} 

Expression of physical, radiometric, and dosimetric quantities should generally follow standard academic conventions \cite{Seltzer2011} as exemplified in Table \ref{tab1}, as well as the International System of Units (SI) \cite{BIPM-brochure,Thompson2008}. 
For example, the SI unit \si{\cm^3} should be used for cubic centimeter, instead of the customary abbreviation cc.
Particles, elements, isotopes, ions, and chemicals may be  written using their symbols and formulas: \gamma, e$^-$, \pi$^-$, p, n, H, He, $^{10}$B, $^{60}$Co,  C$^{6+}$, OH$^-$, H$_2$O, C$_2$H$_5$OH, etc.
Mathematical notation should follow the standard typesetting rules \cite{IUPAC-fonts,NIST-typefaces}:
\begin{itemize}
\item{Italic for single letters that denote constants, scalar variables, and indexes: $c$, $S$, $i$, etc.}
\item{Bold italic for vectors, tensors, and matrices: $\boldsymbol{v}$, $\boldsymbol{T}$, $\boldsymbol{M}$, etc.}
\item{Roman for subscript labels, numerals, units, operators, special math constants, and common functions: $N_\mathrm{A} = \SI{6.02214076e23}{\mol^{-1}}$, \Delta\ (difference), $\mathrm{d}$ (derivative), \pi\ (the circle ratio), $\mathrm{i}=\sqrt{-1}$, $\mathrm{e}$ or $\exp$, $\lg = \log_{10}$, $\ln=\log_\mathrm{e}$, $\lim$, $\max$, $\min$, $\sin$, $\cos$, $\tan$, etc.}
\end{itemize}

\begin{table}
\caption{Standard symbols and units for commonly used physical, radiometric, and dosimetric quantities \cite{Seltzer2011}} 
\label{tab1}
\begin{tabular}{lll}
\toprule
Quantity &Symbol & Units \\
\midrule
length & $l$ & m \\
mass & $m$ & kg, u,$^a$ MeV/$c^2$ $^b$ \\
time & $t$ & s, min, h, d \\
amount of substance & $n$ & mol \\
electric charge & $q$ & C, $e$ $^c$ \\
energy & $E$ & J, eV \\
volume & $V$ & L, \si{\cm^3} \\
density & $\rho$ & \si{\g/\cm^3} \\
cross section & $\sigma$ & \si{\m^2}, b $^d$ \\
lin. attenuation coeff. & $\mu$ & \si{\m^{-1}} \\
lin. stopping power & $S$ & \si{\J/\m}, \si{\keV/\um} \\
particle number & $N$ & 1 \\
flux & $\dot{N}$ & \si{\s^{-1}} \\
fluence & $\varPhi$ & \si{\m^{-2}} \\
dose & $D$ & Gy \\
half life & $T_{1/2}$ & s, min, h, d \\ 
activity & $A$ & Bq \\
\botrule
\end{tabular}
$^a$ The unified atomic mass unit (u or Da) \\ = \SI{1.66053906660e-27}{\kg} = \SI{931.49410242}{\MeV}/$c^2$. \\
$^b$ The speed of light in vacuum $c$ = \SI{299792487}{\m/\s}. \\
$^c$ The elementary charge $e$ = \SI{1.602176634e-19}{\C}. \\
$^d$ 1 barn (b) = \SI{e-28}{\m^2}.
\end{table}

Equations can be either \emph{inline}, such as $E = m c^2$, or \emph{display} between lines, such as
\begin{equation}
E = m c^2, \label{eq1}
\end{equation}
where $E$ is the energy, $m$ is the mass, and $c$ is the photon speed.
Display equations are also inserted into sentences with punctuation and are sequentially numbered for referencing. 
Equation~(\ref{eq1}) to refer to the equation at the beginning of a sentence is otherwise abbreviated as Eq.~(\ref{eq1}). 

\subsection{Tables and figures}

Tables and figures should only be used to present essential information clearly and efficiently.
They typically take the form of a declarative or comparative list for definite parameters; a diagram or drawing for methodology; a chart, graph, or plot for data profiling; and a photograph for visual confirmation.

\begin{figure}
\includegraphics[width=\linewidth]{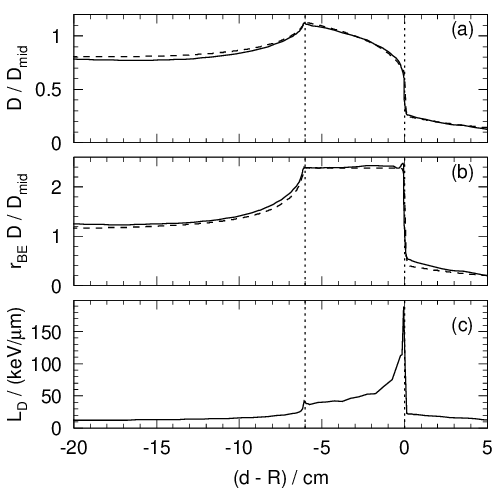}
\caption{\textbf{a} Physical dose $D$ and \textbf{b} RBE-weighted dose relative to mid-SOBP physical dose $r_\mathrm{BE} D$, and \textbf{c} LET as functions of range-subtracted depth $(d-R)$ in water, registered in the HIPLAN (solid lines) and XiO-N (dashed lines) systems for a carbon-ion beam of $E/m$ = 350 MeV/u and range modulation $M$ = 6 cm, with SOBP depth boundaries (vertical dotted lines). \copyright~2018 JSRT \& JSMP. First published in \citet{Kanematsu2018}}
\label{fig1}
\end{figure}

Tables and figures must be referenced in the main text using sequentially numbered labels, such as Table~\ref{tab1} and Fig.~\ref{fig1}, where their content should be explained to readers.
Subfigures in a figure should be labeled using letters \textbf{a}, \textbf{b}, \textbf{c}, and so on.
Figure~\ref{fig1} illustrates an example.

Captions should state the objects and conditions applied, and appropriately identify the elements tabulated or drawn in there.
To include non-original data, the original source must be cited at the end of the caption.
To reuse copyrighted content, the authors must have the right or permission, and justify their use in the caption as agreed with the copyright holder.
Figure~\ref{fig1} also demonstrates such a case.

Tables and figures may be included in the manuscript, preferably near where they are first referenced.
Common graphic formats, typically EPS for vector graphics, PNG for raster graphics, JPEG for photographs, and PDF for all purposes, should be used for figure files submitted with the manuscript to the journal.

\section{Results}

The results section should present new data obtained sequentially with the procedure described in the materials-and-methods section.
They should normally be summarized or visualized in tables or figures, and their textual explanation should be limited to direct observation and natural interpretation relevant to the research objectives.

\section{Discussion}

The discussion section should primarily evaluate and generalize the development, results, and findings presented in the previous sections. 
The significance and impact may be discussed in comparison or in combination with other research or practice as appropriate. 
Any unexpected results should be investigated to see if they can be explained.
Limitations of the current research and anticipated future directions may also be announced.

\section{Conclusions}

The conclusions section should summarize how the research objectives were met and highlight new findings in a quantitative and objective manner.
These conclusions should be written in short sentences and ordered by importance.

\backmatter

\section*{Acknowledgments}

In cases of notable contributions by individuals or parties who do not meet the authorship criteria \cite{ICMJE-recommendations}, the contributors may be acknowledged here with their consent to publication of their names.
In this case, the author thanks Ms. Yuko Matsumoto, Springer Nature Japan, for detailed information on journal-specific requirements and styles.

\section*{Supplementary information}
In certain cases, authors may submit various electronic files with the manuscript.
Supplementary materials for publication, typically data to be shared, must be referenced in the main text with a series of specific labels Online Resource 1, Online Resource 2, and so on.
Their captions must be listed here to concisely describe the file content.
The publisher will make the files available to readers.

\section*{Declarations}

\bmhead{Funding}
Funding sources may be listed according to sponsor requirements. 
In this case, this work was supported by JSPS KAKENHI Grant Number JP20K08013. 

\bmhead{Conflict of interest (or competing interests)}
Any fact that could be considered a potential source of publication bias must be disclosed or affirmatively denied. 
In this case, the author has no conflicts of interest regarding the research, authorship, or publication of this work.

\bmhead{Ethics approval}
The status of the ethics approval or the reasons for the exemption must be stated. 
In this case, the ethics approval was waived because no human or animal participants, specimens, or data were involved in this work.

\bmhead{Informed consent}
For research involving humans, the status of their consent to participation and publication should be stated.

\bmhead{Data availability}
It must be stated whether or not the data used in the paper are available to readers, and if so, how they can be accessed.
In this case, the original \LaTeX\ source files of this guide are openly available on \emph{arXiv} at \url{https://arxiv.org/src/2403.02888}.

\bmhead{Other declarations}
Authors may declare any fact related to the work: code availability, author contributions, etc. 
They may also disclose relevant non-interfering preliminary publications: meeting abstracts, electronic posters, preprints, etc.

\bibliography{rpt-bibliography}

\begin{appendices}

\section{Usage of appendices}\label{appA}

Appendices are a part of the publication to provide the details that may be skipped for the readers to comprehend the research, but are necessary for the completeness.
They must be referenced in the main text with a series of specific labels Appendix A, Appendix B, and so on.

Equations, tables, and figures may be used in the same way as in the main text, but with a new series of labels, for example,
\begin{equation}
-\frac{\mathrm{d}N}{\mathrm{d}t} = \lambda N, \label{eqA1}
\end{equation}
where $\lambda = (\ln 2)/T_{1/2}$ is the decay constant. 
References may also be used in the same way as in the main text \cite{NLM-style}.

\section{Submission files, checklist, and progression}\label{appB}
\setcounter{table}{0} 

Authors must submit their manuscript to the editorial office of \emph{Radiological Physics and Technology} according to the submission guidelines \cite{RPT-guidelines}, using the online submission-and-review system Editorial Manager (Aries Systems Corporation, North Andover, MA, USA) \cite{EM-RPT}.
Table~\ref{tabB1} shows the \LaTeX\ files for this guide, which are an example of submission files.

Supplementary materials for editorial consideration may also be submitted optionally or upon request.
These include a cover letter, conflict-of-interest disclosure, ethics approval, copyright clearance, and a copy of unpublished references. 
The purpose of these additional submission files should be obvious or explained in a metadata field of \emph{author comments} only to editors, or in a cover or revision letter also to reviewers.

Table~\ref{tabB2} is a step-by-step checklist for authors submitting a manuscript.
Table~\ref{tabB3} shows a typical progression of an initial submission, which authors may care about during peer review.

\begin{table}[h]
\caption{\LaTeX\ files for this guide with types to be specified during online submission: M for a main document, F for figures, and S for supporting files including class and style files provided by Springer Nature (SN) \cite{Springer-latex}}
\label{tabB1}
\begin{tabular}{l l c}
\toprule
File name & Content & Type \\
\midrule
rpt-article.tex & main text source & M \\
rpt-bibliography.bib & references source & S \\
figure1.eps & graphic image & F \\
sn-jnl.cls & SN journal class & S \\
sn-vancouver.bst & Vancouver style & S \\ 
\botrule
\end{tabular}
\end{table}

\begin{table*}[p]
\caption{Checklist for manuscript submission to \emph{Radiological Physics and Technology} according to the submission guidelines, using the online submission-and-review system Editorial Manager (EM) \cite{RPT-guidelines, EM-RPT}}
\label{tabB2}
\renewcommand{\arraystretch}{1.5}
\begin{tabular}{c p{0.2 \linewidth} p{0.624 \linewidth} c}
\toprule
No. & Task name & Task description & \textcircled{$\checkmark$} \\
\midrule
1 & \raggedright{Manuscript adaptation} & Carefully adapt the manuscript to the submission guidelines. & \textcircled{ } \\
2 & \raggedright{Language editing} & Optimize the structure and thoroughly edit the language by authors themselves and by any means available. & \textcircled{ } \\
3 & \raggedright{Coauthor and contributor approval} & Send the completed manuscript to all coauthors and acknowledged contributors, and obtain their consent to submission. & \textcircled{ } \\
4 & \raggedright{File collection} & Collect all files for submission: a single-column double-spaced line-numbered text with figures, references, and other \LaTeX\ files, as in Table~\ref{tabB1}, and optional cover letter and other supplementary files. & \textcircled{ } \\
5 & \raggedright{Author comments drafting} & Write down notes and requests to editors to be included in a metadata field of \emph{author comments}: usage of supplementary files, reviewer suggestions, any preferences, priorities, concerns, etc. & \textcircled{ } \\
6 & \raggedright{Novel scientific points drafting} & Itemize three to five catchphrases for the research to appeal to the reviewers within 100 characters ($\approx$ 20 words) each, typically what is new, original, and significant. & \textcircled{ } \\
7 & \raggedright{Online submission} & Log in to EM as an author, upload the collected files, fill in all mandatory metadata fields, and build a PDF manuscript for review. Carefully verify and approve the manuscript. Then, it is sent to the editorial office as a new submission, and an inquiry e-mail letter is sent to coauthors. & \textcircled{ } \\
9 & \raggedright{Coauthor confirmation} & Ask all coauthors to click on an electronic link in the letter to confirm their authorship. Check the \emph{author status} with EM until all have confirmed, or resend the letter to those who haven't and remind them again. & \textcircled{ } \\
10 & \raggedright{Editorial receipt verification} & Verify with EM that the editorial office has received the submission and sent it to editors for peer review. Incomplete submissions will be sent back to the authors. & \textcircled{ } \\
11 & \raggedright{Peer review monitoring} & Monitor the progression of \emph{editorial status} as described in Table~\ref{tabB3} regularly with EM. If it takes more than two months, consider contacting the editorial office for details. & \textcircled{ } \\
\botrule
\end{tabular}
\end{table*}

\begin{table*}
\caption{Typical progression of \emph{editorial status} for an initial submission to \emph{Radiological Physics and Technology}. The actors include the journal editorial office assistant (JEOA), editor-in-chief (EIC), deputy editor (DE), associate editor (AE), authors, and reviewers} 
\label{tabB3}
\renewcommand{\arraystretch}{1.5}
\begin{tabular}{p{0.12 \linewidth} p{0.09 \linewidth} p{0.69 \linewidth}}
\toprule
Status & Actor & Ongoing or expected action \\
\midrule
\raggedright{Submitted to Journal} & JEOA & {Technically verifying the submitted files, metadata, and \emph{author status}, and applying an initial quality check including plagiarism detection.} \\
\raggedright{Sent Back to Author} & authors & {Redoing the manuscript submission that was found to be incomplete.} \\
\raggedright{With Editor} & EIC/DE & {Assessing the submission for the journal scope, or considering possible transfer to another journal or rejection.} \\
\raggedright{Editor Invited} & DE & {Contacting an AE as a potential editor to handle the peer review process.} \\
\raggedright{With Editor} & AE & {Analyzing the manuscript and identifying the expertise required for reviewers.} \\
\raggedright{Reviewer Invited} & AE & {Contacting experts as potential reviewers.} \\
\raggedright{Under Review} & reviewers & {Reviewing the manuscript to report review comments independently.} \\
\raggedright{Required Reviews Completed} & AE & {Analyzing the collected review comments with necessary edits to make an editorial recommendation.} \\
\raggedright{Decision in Process} & DE/EIC & {Confirming the recommendation and finalizing the review comments to make an editorial decision.} \\
\botrule
\end{tabular}
\end{table*}

\section{Peer review, revision, and final decision}\label{appC}
\setcounter{table}{0} 

Peer review is inherently a collaborative process among authors, reviewers, and editors, who all wish to advance the scientific field they share.
The reviewers are generally experts equivalent to the authors, which allows them to evaluate the scientific content and to help authors increase its value. 

When the authors receive a decision letter for revision, they should use the review comments to make the paper clearer, free of errors or oversights, and more appealing to readers.
In general, reviewers' misunderstandings or questions indicate a need for clarification of the manuscript.
In a revision letter, the authors must copy all the comments and insert responses between the lines to state their interpretations and how they changed the manuscript on a point-by-point basis.
In cases where the authors have decided not to change the manuscript as requested, they must provide a logical rebuttal to justify their decision.
Nontrivial voluntary changes must also be justified.
The journal additionally requires a marked manuscript with highlighted changes to be uploaded with a clean manuscript and a revision letter.
If necessary, a \href{https://resource-cms.springernature.com/springer-cms/rest/v1/content/7454878/data/v5}{change of authorship request form} provided by the publisher may be used and included in a revision.

In the case of rejection, the authors should try to understand the reasons from the review comments.
If the editors see potential for substantial rework that could lead to publication, they will encourage resubmission for another evaluation by the same reviewers.
Otherwise, they will not receive essentially the same work, and a remaining option may be resubmission to another journal.
If the authors are not convinced with the decision, they may appeal to the editor-in-chief with reasons for reconsideration.
Generally, all such communications should be made using Editorial Manager or by e-mail through the editorial office.

Original papers must have sufficient scientific value in terms of relevant new knowledge.
Opinions about value may differ among researchers, and a consensus of a few can never be absolute, but only a fair decision in the peer review system.
The results of peer review should not be taken too seriously, but authors, reviewers, and editors are expected to learn and grow as researchers through these processes.

\end{appendices}

\end{document}